\newcommand{\CLASSINPUTtoptextmargin}{2cm}
\newcommand{\CLASSINPUTbottomtextmargin}{2cm}
\documentclass[conference,1opt,final,a4paper]{IEEEtran}
\usepackage{amsmath}
\usepackage{amssymb}
\usepackage{xcolor}
\usepackage{amsthm}
\usepackage{mathtools}
\usepackage{siunitx}
\usepackage{cuted}
\usepackage{graphicx}
\usepackage{subcaption}
\usepackage{epstopdf}

\usepackage{tabularx}
\usepackage{multirow}
\usepackage{booktabs}
\usepackage{diagbox}

\usepackage[norelsize]{algorithm2e}
\makeatletter
\newcommand{\removelatexerror} {\let\@latex@error\@gobble}
\makeatother

\usepackage{tcolorbox}
\tcbuselibrary{many}

\newcommand{\superscript}[1]{^{\text{#1}}}
\newcommand{\subscript}[1]{_{\text{#1}}}

\newcommand\blfootnote[1]{%
	\begingroup
	\renewcommand\thefootnote{}\footnote{#1}%
	\addtocounter{footnote}{-1}%
	\endgroup
}

\newtcbtheorem[]{alg}{Algorithm}{fonttitle=\bfseries}{alg}

\IEEEoverridecommandlockouts

\begin{document}

\title{Optimal Blocklength Allocation towards Reduced Age of Information in Wireless Sensor Networks}


\author{
	\IEEEauthorblockN{Bin~Han\IEEEauthorrefmark{1}, Yao~Zhu\IEEEauthorrefmark{2}, Zhiyuan~Jiang\IEEEauthorrefmark{3}, Yulin~Hu\IEEEauthorrefmark{2}, and Hans~D.~Schotten\IEEEauthorrefmark{1}\IEEEauthorrefmark{4}}
	\IEEEauthorblockA{
		\IEEEauthorrefmark{1}Technische Universit\"at Kaiserslautern, \IEEEauthorrefmark{2}RWTH Aachen University, \IEEEauthorrefmark{3}Shanghai University, 
		\IEEEauthorrefmark{4}DFKI
	}%
}


\maketitle

\begin{abstract}
The freshness or timeliness of data at server is a significant key performance indicator of sensor networks, especially in tolerance critical applications such as factory automation. As an effective and intuitive measurement to data timeliness, the metric of Age of Information (AoI) has attracted an intensive recent interest of research. This paper initiates a study on the AoI of wireless sensor networks working in the finite blocklength (FBL) regime as a resource allocation problem, and proposes to minimize the long-term discounted system AoI as a Markov decision process (MDP). The proposed method with its optimum solved by Reinforced Learning technique is verified by simulations to outperform benchmarks, including the conventional error rate minimizing policy.
\end{abstract}


\IEEEpeerreviewmaketitle

\blfootnote{\copyright 2019 IEEE.  Personal use of this material is permitted. Permission from IEEE must be obtained for all
	other uses, in any current or future media, including reprinting/republishing this material for advertising
	or promotional purposes, creating new collective works, for resale or redistribution to servers or lists, or
	reuse of any copyrighted component of this work in other works. }

\section{Introduction}\label{sec:introduction}
Wireless sensor networks play a key important role in various applications such as environmental monitoring, target tracking, smart grid and factory automation~\cite{Yick_2008}. They are expected to deliver data with satisfactory freshness, which is especially critical in industrial scenarios to ensure a smooth and safe functioning of the system~\cite{Islam_2012}.

To characterize information freshness, recently, the metric of Age of Information (AoI)~\cite{kaul12} has been proposed. AoI denotes the time elapsed since the generation time of the last successfully received status. Basically, AoI is a time metric, however, it is fundamentally different from conventional delay metrics and requires special attention. Most existing works focus on AoI optimization as a scheduling problem in Medium-Access-Control (MAC) layer, while the implication on physical layer design is largely ignored.

When shifting our focus to the bottom of physical layer, we can realize that transmission errors in radio access network can also significantly impact the AoI in sensor networks, in a similar way like they influence the uplink delay. Nevertheless, with its special feature of memory, the AoI will probably exhibit a different behavior, in comparison to the delay metrics, with respect to the transmission error rate, which has not been well studied so far.

In all kinds of wireless networks, physical layer error control universally relies on resource allocation techniques in different dimensions including power, bandwidth and time. In context of uplink data transmission for sensor networks, a flexible and commercially practical solution is to allocate blocklength (time/bandwidth) among sensor devices in a TDMA manner. Especially, when data packet size are limited, the blocklength of the transmission is short, the Shannon capacity does not hold and the data transmission becomes no long arbitrarily reliable.  Recently, this packet error rate (PER) due to short blocklength is characterized in~\cite{Polyanskiy_2010}, in a so-called finite blocklength (FBL) regime. 
Following the FBL model,  a quantitative dependency of the PER on blocklength assigned to every device is obtained, so that a cell-level error minimization is enabled. 

In this work, we investigate the AoI of sensor network systems working in the FBL regime. In a physical layer perspective, we discuss the following resource allocation problem, which yet has not been studied by literature to the best of our knowledge: \emph{How should blocklength be allocated to different sensors, in order to minimize the overall network AoI?}

The remainder of this paper is organized as follows: Sec.~\ref{sec:stoa} reviews the state-of-the-art in both fields of AoI and FBL. Sec.~\ref{sec:methods} models the system AoI as a function of blocklength allocation, and propose two optimization approaches, which minimize the system AoI in short and long terms, respectively. In Sec.~\ref{sec:simulations} we present the procedure, results and analysis of the numerical simulations, through which our proposed methods are evaluated together with benchmarks. After some additional discussions in Sec.~\ref{sec:discussions}, we conclude this paper with Sec.~\ref{sec:conclusion}.


\section{Relevant Studies}\label{sec:stoa}
\subsection{AoI as a Scheduling Problem}
Previous works on AoI mainly focus on the MAC-layer problem where the data sink (server) can be busy or idle to process the data packets generated by sources (sensors). Literature has derived~\cite{huang19} that it helps to reduce AoI by replacing the outdated packets in queue and with the latest one from the same source. Therein, the dropping of data packets is caused by queue congestions. In such context,
it has been derived that both over-high and over-low sampling rates of a sensor lead to an increased expectation of the AoI~\cite{kaul12}. Moreover, in a previous work~\cite{Jiang_2019} we have shown that  the mean uplink AoI at sink (server) in a simple TDMA master-slave system is linear to the number of sources (sensors), if all sources equally share the same frame length. Furthermore, scheduling schemes for AoI optimization in various perspectives have been well studied in the literature~\cite{hsu18,kadota18,jiang19,jiang18_isit,jiang18_itc}.

\subsection{Error Control in FBL Regime}
FBL information theory sets up an exciting binding between the transmission slot length and the transmission error rate~\cite{Polyanskiy_2010}. Furthermore, to enhance the performance, the retransmission mechanisms in FBL regime are also introduced, while aiming at minimizing the energy consumption~\cite{Avranas_2018}. The trade-off between reliability and energy efficiency is also studied in a retransmission-enabled edge computing scenario~\cite{Yao_SPAWC_2019}. Nervertheless, similar to the case of AoI studies, in FBL regime it is never encouraged to retransmit the same package with respect to error probability, either, as literature has shown that the best performance achievable by HARQ, even when neglecting the feedback loss, is equal to the performance of optimal one-shot transmission~\cite{Makki_2014}. In particular, 
the total errors with cooperative nodes can be minimized by a time resource allocation that grants nodes with the blocklengths that leads to the same error rate for every device~\cite{Yao_2019}.

\section{Problems, Models and Approaches}\label{sec:methods}

\subsection{Model Setup}
Now we consider the AoI problem in perspective of PHY-layer transmission errors as follows: Multiple sensors are scheduled to transmit messages to a Multi-Access Edge Computing (MEC) server in a FBL-TDMA manner, where all sensors are synchronous to the same uplink transmission period. Every message generated by every device has the same bit-length. The uplink channels of different sensors are independent from each other. If a message fails to be delivered to the server, no retransmission is scheduled, and the sensor will just transmit the latest message in every period. For simplification we assume that all the sensors are active sources (i.e. they generate timely information at their scheduled transmission time), the server is always idle to process any incoming message, and the uplink channel of every sensor is non-fading additive white Gaussian noise (AWGN) channel with full channel state information (CSI) at the MEC server.

Now consider for the $m\superscript{th}$ sensor in the schedule of an $M$-sensor system. Upon the success/failure of transmission, its AoI at the end of $k\superscript{th}$ period is
\begin{equation}
	A_m(k)=\begin{cases}
		1&\text{success};\\
		A_m(k-1)+1&\text{failure}.
	\end{cases}
\end{equation}
Thus, the expected sum of AoI at the end of $k\superscript{th}$ period is
\begin{equation}
	\begin{split}
		&\mathbb{E}\left\{\lvert\mathbf{A}_k\rvert\right\}=\sum\limits_{m=1}^M\left\{(1-\varepsilon_m)+\varepsilon_m\left[A_m(k-1)+1\right]\right\}\\
		=&M+\sum\limits_{m=1}^M\varepsilon_mA_m(k-1)
	\end{split}
\end{equation}

\subsection{Single-Period AoI Optimization}
	Given the AoI of every sensor at the beginning of $k\superscript{th}$ period, it is a natural idea to attempt minimizing $\lvert\mathbf{A}_k\rvert$ with respect to the blocklengths $\mathbf{n}=[n_1,n_2,\dots,n_M]\in\mathbb{N}^M$:
	\begin{align}
		\min\limits_{\mathbf{n}}\mathbb{E}\left\{\lvert\mathbf{A}_k\rvert\right\}&=M+\min\limits_{\mathbf{n}}\sum\limits_{m=1}^M\varepsilon_m(n_m)A_m(k-1)\label{eq:one-step-aoi-min}\\
		\text{s.t.}\quad&\sum\limits_{m=1}^Mn_m\le\left\lfloor\frac{T}{T\subscript{S}}\right\rfloor\overset{\Delta}{=}N_{\max},\\
		&n_m\ge n_{m,\min},\forall m\in\{1,2,\dots,M\},
	\end{align}  
	where $T\subscript{S}$ is the symbol length and $n_{m,\min}$ is the minimal blocklength required by sensor $m$ w.r.t. the maximal allowed package error rate $\varepsilon_{\max}$. Especially, in the FBL regime according to \cite{Polyanskiy_2010} we have
	\begin{equation}\label{eq:err_rate}
		\varepsilon_m(n_m)\approx Q\left(\sqrt{\frac{n_m}{V_m}}\left(\mathcal{C}_m-\frac{d_m}{n_m}\right)\ln 2\right),
	\end{equation}
	where $\mathcal{C}_m$, $V_m$ and $d_m$ denote the Shannon capacity, channel dispersion and message length of sensor $m$, respectively.
	
	As a reference, the classic FBL problem that aims at the minimization of system PER can be formulated as
	\begin{equation}
		\min\limits_{\mathbf{n}}\sum\limits_{m=1}^M\varepsilon_m\label{eq:min_err_rate},
	\end{equation}
	which differs from \eqref{eq:one-step-aoi-min} by a linear coefficient $A_m(k-1)$ in every term.
	
	Similar to the approach  used in classic FBL problems, here we relax the constraint $\mathbf{n}\in\mathbb{N}^M$ to $\mathbf{n}'\in\left(\mathbb{R}^+\right)^M$. In this case it is trivial to prove that the minimum \eqref{eq:one-step-aoi-min} is achieved when
	\begin{align}
		\sum\limits_{m=1}^M\tilde{n}_m&=N_{\max},\label{eq:relaxed_sum}\\
		\frac{\varepsilon_i(n_i')}{\varepsilon_j(n_j')}&=\frac{A_j(k-1)}{A_i(k-1)}, \forall [i,j]\in\{1,2,\dots,M\}^2\label{eq:relaxed_opt},
	\end{align}
	and $\mathbf{n}\subscript{opt}\in\mathbb{N}^M$ can be then approximated by rounding $\mathbf{n}'\subscript{opt}$.
	
	Remark that due to the non-linearity of \eqref{eq:err_rate}, the relaxed problem (\ref{eq:relaxed_sum},~\ref{eq:relaxed_opt}) is analytically solvable only when $A_i(k-1)=A_j(k-1), \forall[i,j]$. In general cases, we have to rely on traversing over the solution vector space $\mathbb{V}\subset\{1, 2,\dots,N_{\max}\}^M$ to find the approximate optimum:
	
	\begin{equation}
		\mathbf{n}'\subscript{opt}\approx\min\limits_{\mathbf{n}'\in\mathbb{V}'}\sum\limits_{i=1}^M\left(\xi_i-\frac{1}{M}\sum\limits_{j=1}^M\xi_j\right)^2,
	\end{equation}
	where $\xi_i=\varepsilon_i(n_m)A_m(k-1)$.

\subsection{Long-Term Optimization}
Nevertheless, it shall be noted that the single-period AoI minimization discussed above has ignored the future impact of current decision on the system, as AoI is a feature with memory. In long term, this may lead to a convergence at local sub-optimum instead of global optimum, which reduces the optimization gain. To cope with this issue, in this section we consider the long-term AoI optimization. 

To simplify the discussion we consider here a two-sensor case without fading where $M=2$, both $[\mathcal{C}_1,\mathcal{C}_2]$ and $[V_1,V_2]$ are constant. We also reasonably assume the system to be initialized with a certain AoI state $\mathbf{A}_0$.

Consider a consistent policy $\mathcal{P}$:
\begin{equation}
	\left(\mathbb{N}^+\right)^2\overset{\mathcal{P}}{\to}\{n_{1,\min},1,\dots,(N_{\max}-n_{2,\min})\},
\end{equation}
which maps the AoI state at the beginning of $k\superscript{th}$ period $\mathbf{A}_{k-1}$ to the time allocation $[n_1(k), N_{\max}-n_1(k)]$ , so we can rewrite $n_1(k)$ as $n_1(\mathbf{A}_{k-1})$, $\varepsilon_1(n_1)$ as $\varepsilon_1(\mathbf{A}_{k-1})$, and $\varepsilon_2(n_2)$ as $\varepsilon_2(\mathbf{A}_{k-1})$. The state transition probability $P(\mathbf{A}_{k}\vert\mathbf{A}_{k-1})$ relies only on $\mathbf{A}_{k-1}$ therefore :

\begin{strip}
	\begin{tcolorbox}
		\begin{equation}
		P(\mathbf{A}_{k}\vert\mathbf{A}_{k-1})=
		\begin{cases}
		\varepsilon_1(\mathbf{A}_{k-1})\varepsilon_2(\mathbf{A}_{k-1})&\mathbf{A}_k=[A_1(k-1)+1,A_2(k-1)+1]\\
		\varepsilon_1(\mathbf{A}_{k-1})[1-\varepsilon_2(\mathbf{A}_{k-1})]&\mathbf{A}_k=[A_1(k-1)+1,1]\\
		[1-\varepsilon_1(\mathbf{A}_{k-1})]\varepsilon_2(\mathbf{A}_{k-1})&\mathbf{A}_k=[1,A_2(k-1)+1]\\
		[1-\varepsilon_1(\mathbf{A}_{k-1})][1-\varepsilon_2(\mathbf{A}_{k-1})]&\mathbf{A}_k=[1,1]\\
		0&\text{otherwise}
		\end{cases}\label{eq:transition}
		\end{equation}
	\end{tcolorbox}
\end{strip}

Thus, this becomes an infinite-state Markov Decision Process (MDP) that can be optimized to minimize the long-term discounted AoI:
\begin{equation}
	\min\limits_{\mathcal{P}}\mathbb{E}\left\{\left.\sum\limits_{k=1}^{+\infty}\gamma^{k-1}\lvert\mathbf{A}_k\rvert~\right\vert~\mathbf{A}_0\right\},\label{eq:mdp_infinite}
\end{equation}
where $\gamma\in(0,1)$ is the discount factor.

As the weight $\gamma^{k-1}$ falls exponentially w.r.t. $k$, here we can set a finite yet sufficient term length $K$ that $\gamma^K\approx 0$, in order to reduce the computational effort. Furthermore, it shall be noted that \eqref{eq:transition} implies
\begin{equation}
	\text{Prob}\{A_m(k)=a\}\sim\prod\limits_{i=1}^a\varepsilon_m(\mathbf{A}_{i-1}),\quad m\in\{1,2\},
\end{equation}
where $\varepsilon_m(\mathbf{A_{i-1}})\in(0,\varepsilon_{\max})$ always hold. Practically we can find some finite $A_{\max}$ that $\text{Prob}\{A_m(k)>A_{\max}\}\approx 0$. Thus, the infinite-state MDP optimization problem over infinite time \eqref{eq:mdp_infinite} can be approximately degraded to a finite-state MDP optimization problem over $K$ periods:
\begin{gather}
	\min\limits_{\mathcal{P}'}\mathbb{E}\left\{\left.\sum\limits_{k=1}^{K}\gamma^{k-1}\lvert\mathbf{A}_k\rvert~\right\vert~\mathbf{A}_0\right\},\label{eq:mdp_finite}\\
	\{1,2\dots A_{\max}\}^2\overset{\mathcal{P'}}{\to}\{n_{1,\min},1,\dots,(N_{\max}-n_{2,\min})\}.
\end{gather}

It is common to solve problems such as \eqref{eq:mdp_finite} with Reinforced Learning (RL) approaches, e.g. the well-known Q-Learning algorithm. In this algorithm, a so-called Q-matrix $\mathbf{Q}_{I\times J}$ is constructed to represent the expected discounted rewards of different actions (time allocations) in all possible system states (AoI), where $I=A_{\max}^2$ is the number of possible states, and $J=N_{\max}-n_{2,\min}-n_{1,\min}$ is the number of valid actions. Two mappings $\{1,2\dots I\}\overset{\Theta}{\to}\{1,2\dots A_{\max}\}^2$ and  $\{1,2\dots J\}\overset{\Omega}{\to}\{n_{1,\min},n_{1,\min}+1\dots N_{\max}-n_{2,\min}\}$ are defined here for convenience of the matrix index notation. The offline learning process to train the Q-matrix is briefly described by Algorithm~\ref{alg:rl_solver}. Remark that the notations $\mathbf{Q}$ and $Q_{i,j}$ here are irrelevant to the $Q$ function in Eq.~\eqref{eq:err_rate}.
\begin{alg}{Q-Learning-based MDP solver}{}{}
	\removelatexerror
	\begin{algorithm}[H]
		\label{alg:rl_solver}
		\footnotesize
		Specification: $l_{\max}, \epsilon_{\min},\gamma$\;
		Initialization: $l=1; \epsilon=0; Q_{i,j}=-\infty,\forall{i,j}$\;
		\For(\hfill\emph{Iterative updating}){$l\le l_{\max}$}{
			\For(\hfill\emph{Traversing over all states}){$1\le i\le I$}{
				$Q\subscript{old}\gets\max\limits_{1\le j\le J}Q_{i,j}$\;
				$Q_{\max}\gets-\infty$\;
				\For(\hfill\emph{Traversing over all actions}){$1\le j\le J$}{
					$\mathbf{A}_{k-1}\gets\Theta(i)$\;
					$n_1\gets\Omega(j)$\;
					$r\gets-\sum\limits_{\mathbf{A}_k}\vert\mathbf{A}_k\vert P(\mathbf{A}_k\vert\mathbf{A}_{k-1}, n_{1})$\;
					$Q_{i,j}\gets r+\gamma\max\limits_{1\le j'\le J}Q_{\Theta(i),j'}$\;
					$Q_{\max}\gets\max\{Q_{\max},Q_{i,j}\}$
				}
				$\epsilon\gets\max\{\epsilon,\vert Q_{\max}-Q\subscript{old}\vert\}$
			}
		\If(\hfill\emph{Convergence}){$\epsilon\le\epsilon_{\min}$}{\textbf{Break}}
		}
	\end{algorithm}
\end{alg}
Having the training process accomplished, the optimal policy $\mathcal{P}'\subscript{opt}$ is obtained by
\begin{equation}
	\mathcal{P}'\subscript{opt}(\mathbf{A}_k)=\Omega^{-1}\left(\arg\max\limits_{1\le j\le J}Q_{\Theta^{-1}(\mathbf{A}_k),j}\right).\label{eq:mdp_opt_policy}
\end{equation}

\section{Simulations}\label{sec:simulations}
\subsection{Simulation Setup}
To verify the AoI performance of the proposed methods, we simulate a two-sensor system working over non-fading AWGN channels, where each sensor is granted a bandwidth normalized to unity $B=1$. In each period, the two sensors share a total blocklength of $N_{\max}=500$ times symbol length, each sensor attempting to transmit a packet of $d=16$ bits to the MEC server in uplink, with a maximal allowed PER $\varepsilon_{\max}=0.1$. The sensor AoI at the MEC server is initialized at $\mathbf{A}_0=[1,1]$. We consider four scenarios with different SNR specifications, as listed in Table \ref{tab:snr_specs}.

Under every specification we conduct a $500$-time Monte-Carlo experiment, where in every individual test we simulate the system over $K=500$ periods with both the single-period AoI optimal policy (which is obtained through an exhaustive traversing over the solution space) and the long-term optimal policy. The system AoI state $\mathbf{A}$ is recorded every period. For the long-term optimum, we model the MDP with maximal sensor AoI $A_{\max}=8$, discount factor $\gamma=0.9$, and set the convergence conditions of the Q-Learning-based MDP solver to $l_{\max}=100, \epsilon_{\min}=1~\mathrm{e}-5$.

As a benchmark we also test the standard FBL resource allocation policy~\cite{Yao_2019} that minimizes the overall uplink transmission PER in every period by solving \eqref{eq:min_err_rate}, as well as the simple TDMA scheme that uniformly allocate the time resource to both sensors, i.e. $n_1\equiv n_2$.
\begin{table}[!tbp]
	\centering
	\caption{Uplink SNRs in different scenarios}
	\begin{tabular}{l|c|c|c|c}
		\toprule[1.5px]
										&\textbf{Scenario 1}&\textbf{Scenario 2}&\textbf{Scenario 3}&\textbf{Scenario 4}\\\midrule[1px]
		\textbf{Sensor 1}	&\SI{-13}{\dB}			&\SI{-13}{\dB}			&\SI{-10}{\dB}				&\SI{-8}{\dB}\\
		\textbf{Sensor 2}	&\SI{-6}{\dB}			&\SI{-3}{\dB}				&\SI{-3}{\dB}				&\SI{-8}{\dB}\\\bottomrule[1.5px]
	\end{tabular}
	\label{tab:snr_specs}
\end{table}

\subsection{Results and Analysis}
\begin{figure}[!bp]
	\centering
	\includegraphics[width=\linewidth]{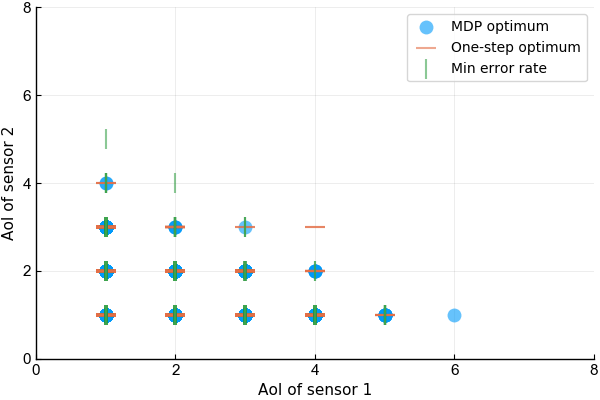}
	\caption{Scatter plot of the system AoI state in Scenario 1}
	\label{fig:scatter_state}
\end{figure}

To ensure $A_{\max}=8$ is sufficient for $\text{Prob}(A_m(k)>A_{\max})\approx0,\forall k\le K$  to hold, we examine the scatter plots of $\mathbf{A}$ generated by the tested policies. For example, with the simulation environments specified to Scenario 1, the system AoI states under different policies are shown in Fig. \ref{fig:scatter_state}, where it can be observed that neither $A_1(k)$ nor $A_2(k)$ has ever exceeded 6 over the simulation trace. This implies the validness of our assumption and therefore it holds $\mathcal{P}'\subscript{opt}=\mathcal{P}\subscript{opt}$. The results for Scenarios 2--4 are similar.
\begin{figure*}[!tbp]
	\centering
	\begin{subfigure}{.45\linewidth}
		\centering
		\includegraphics[width=\linewidth]{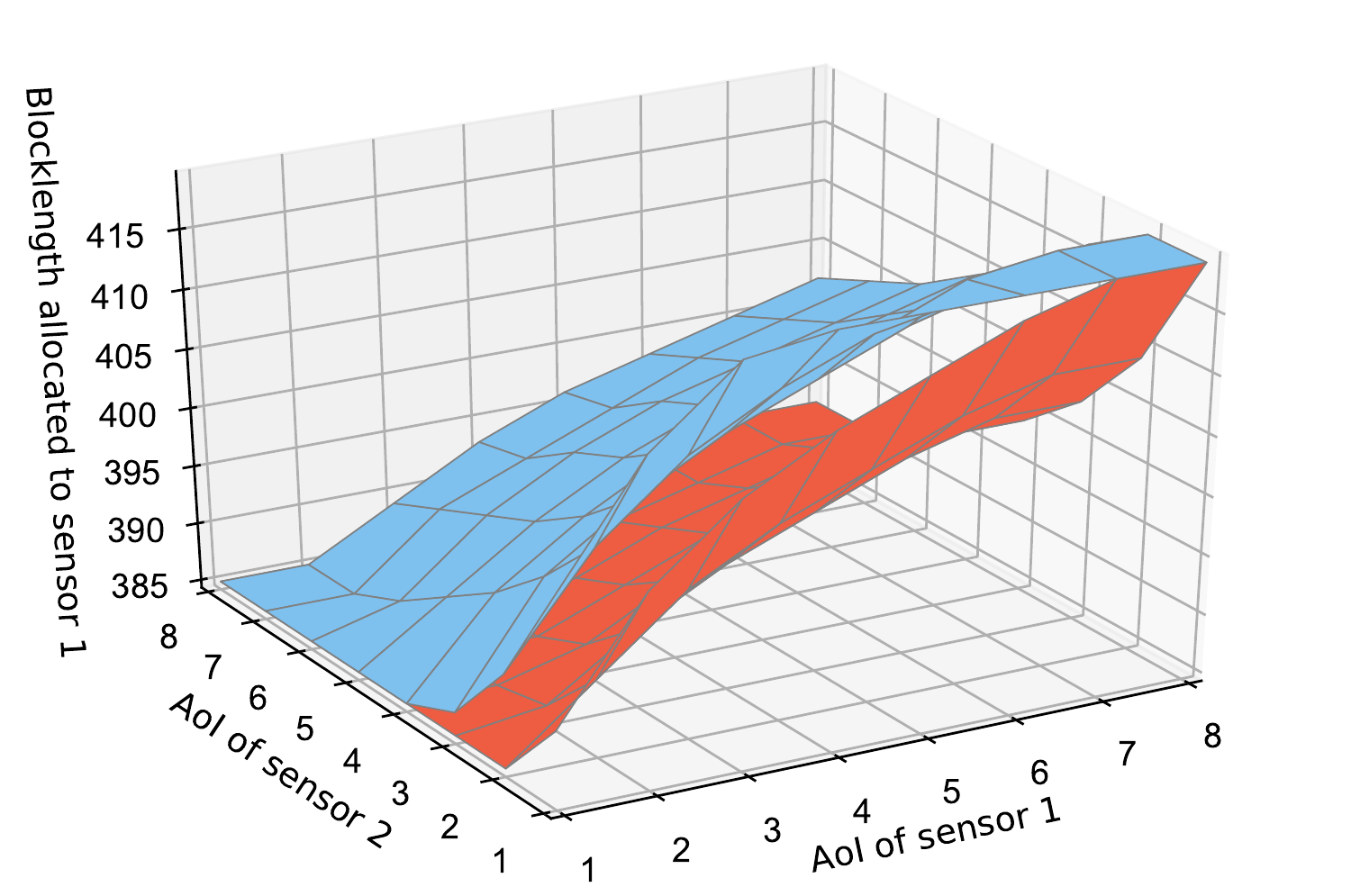}
		\caption{}
		\label{fig:policies_1}
	\end{subfigure}
	\begin{subfigure}{.45\linewidth}
		\centering
		\includegraphics[width=\linewidth]{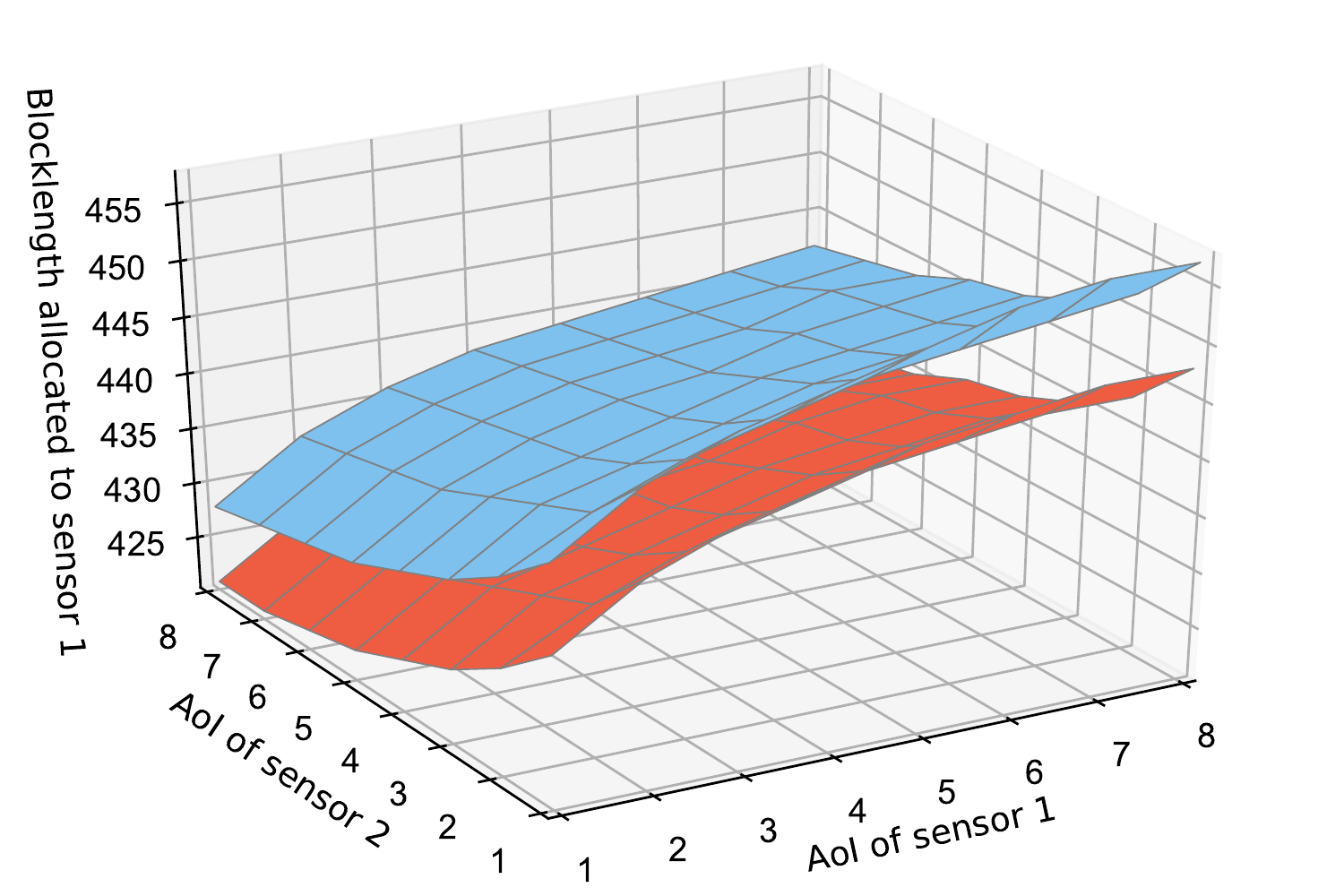}
		\caption{}
		\label{fig:policies_2}
	\end{subfigure}
	\begin{subfigure}{.45\linewidth}
		\centering
		\includegraphics[width=\linewidth]{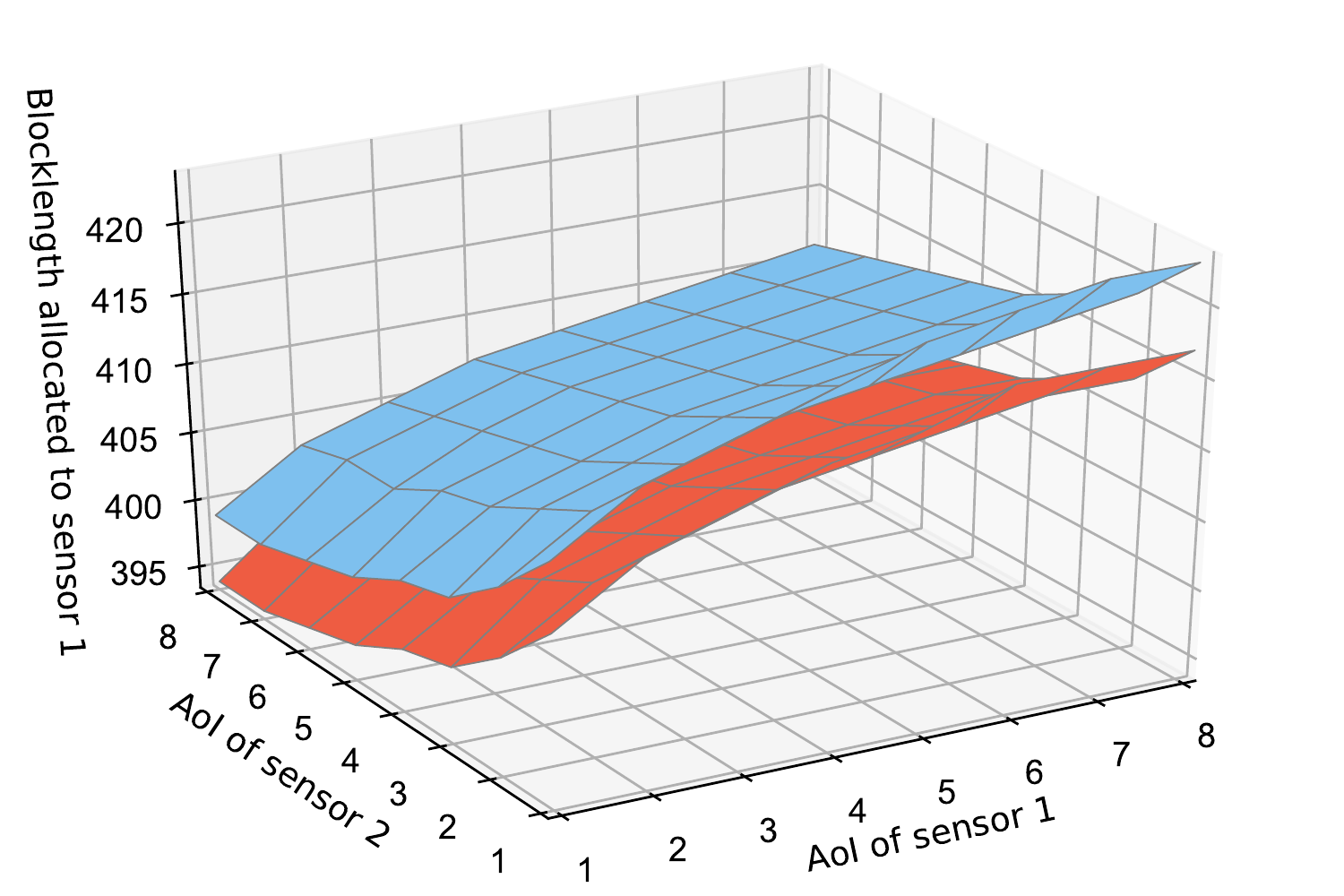}
		\caption{}
		\label{fig:policies_3}
	\end{subfigure}
	\begin{subfigure}{.45\linewidth}
		\centering
		\includegraphics[width=\linewidth]{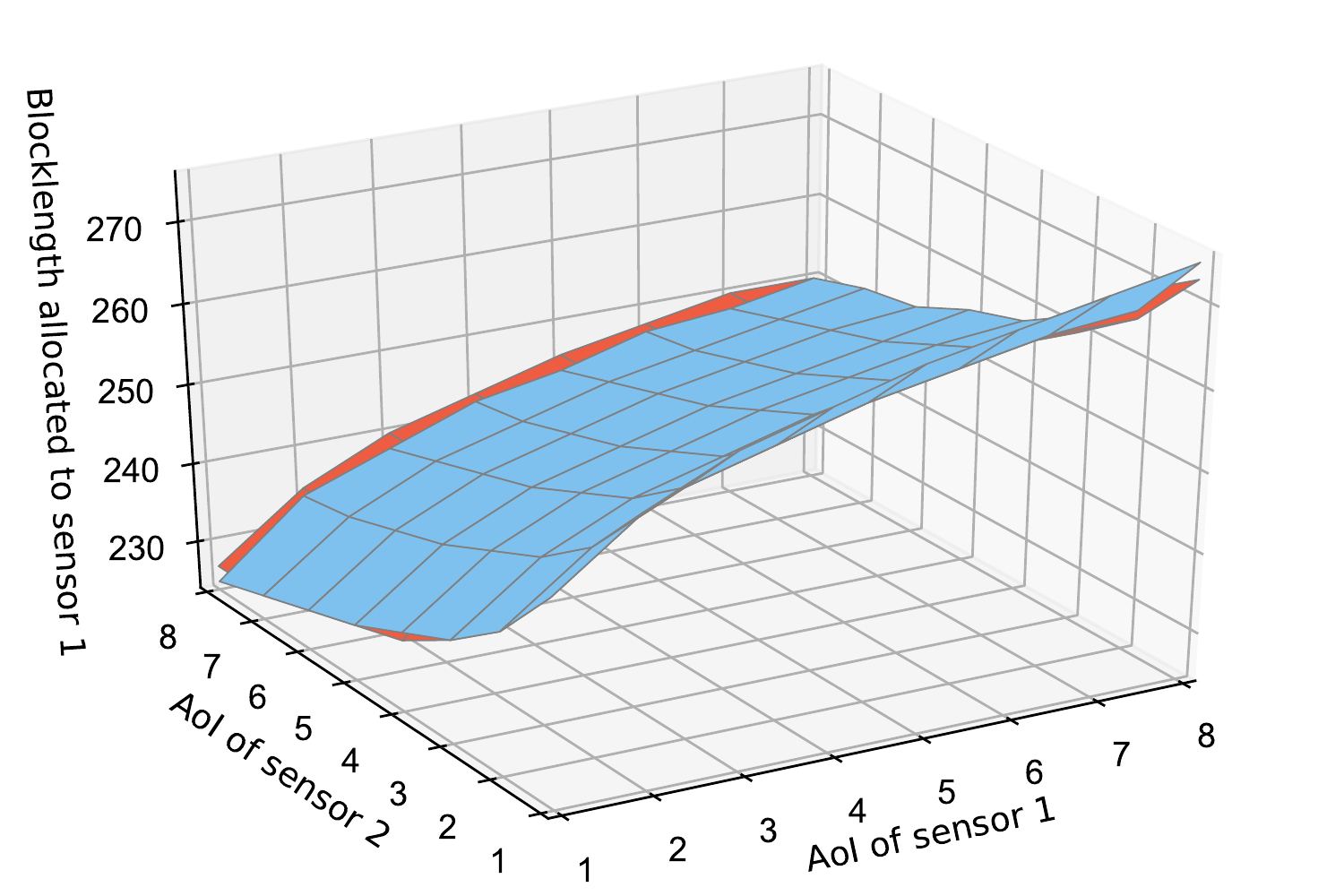}
		\caption{}
		\label{fig:policies_4}
	\end{subfigure}
	\caption{The MDP $n_1\superscript{lt}$ (red) and one-step optimum $n_1\superscript{os}$ (blue) of blocklength assigned to sensor 1 as functions of $\mathbf{A}$, (a)--(d) for Scenarios 1--4, respectively.}
	\label{fig:policies}
\end{figure*}

Next, we compare the policies of MDP optimum and one-step optimum, as visualized in Fig.~\ref{fig:policies}, which shows the blocklengths assigned to sensor 1 in different scenarios, by the long-term optimal policy ($n_1\superscript{lt}$, red) and the one-step optimal policy ($n_1\superscript{os}$, blue), respectively.  In cases of different channel conditions for the two sensors (Scenarios 1--3), it can be observed that in comparison to the one-step optimum, the long-term optimal policy obtained by solving the MDP generally tends to reserve more blocklength for the sensor with better channel (sensor 2). When both sensors have the same SNR, this difference becomes negligible. Remark that both benchmarks, i.e. the PER minimizing strategy and the equal blocklength allocation policy, are independent from the current system AoI $\mathbf{A}$, as listed in Tab.~\ref{tab:benchmarks}.

Then we evaluate the performances of both policies together with the benchmarks. First, for every individual 500-period test we calculate the long-term discounted AoI
\begin{equation}
D=\sum\limits_{k=1}^K\gamma^{k-1}\vert\mathbf{A}_k\vert,
\end{equation}
then for every set of Monte-Carlo test we investigate the average discounted AoI of 500 experiments, subtracting from it the lower bound $D\subscript{lower}=20$ which is the value of $D$ when $\mathbf{A}_k\equiv[1,1]$:
\begin{equation}
\Delta D=D-D\subscript{lower}=D-\sum\limits_{k=1}^K2\gamma^{k-1}=D-20.
\end{equation}
We also calculate the variance of $D$ among 500 experiments. The long-term performances are listed in Tab.~\ref{tab:longterm_performance}.

In addition we study the undiscounted AoI performance as well. For every policy in every individual scenario, we track the instantaneous system AoI ${\vert\mathbf{A}\vert}$ through $500\times 500$ simulated periods,  then calculate its mean value and variance, which are listed in Tab.~\ref{tab:instant_performance}. Similarly, the lower bound is subtracted from the mean value for a more intuitive comparison:
\begin{equation}
\Delta\overline{\vert\mathbf{A}\vert}=\overline{\vert\mathbf{A}\vert}-{\vert\mathbf{A}\vert}\subscript{lower}=\overline{\vert\mathbf{A}\vert}-2.
\end{equation}

From the tables it can be clearly concluded that the MDP approach outperforms all other three methods in both discounted and undiscounted AoI performances by providing lower and stabler system AoI, especially when the channel is harsh and non-uniform for different sensors. The na\"ive TDMA approach with equal blocklength allocation fails to deliver satisfactory performance in most scenarios.  It is interesting to observe that the AoI performance of one-step AoI minimization approach hardly differs from the PER minimization method, which implies that local optimum to minimize AoI in every individual period ignores the impact of current decision in future system states, which strongly reduces the overall performance gain it brings to the system.
\begin{table}[!tbp]
	\centering
	\caption{Optimal blocklength $n_1$ assigned to sensor 1 with different benchmark policies and reference scenarios}
	\begin{tabular}{c|c|c|c|c}
		\toprule[2px]
		\diagbox{\textbf{Policy}}{\textbf{Scenario}}&\textbf{1}&\textbf{2}&\textbf{3}&\textbf{4}\\\midrule[1px]
		\textbf{Min. PER}&402&442&411&250\\\hline
		\textbf{Uniform allocation}&\multicolumn{4}{c}{250}\\
		\bottomrule[2px]
	\end{tabular}
	\label{tab:benchmarks}
\end{table}

\begin{table}[!tbp]
	\centering
	\caption{Long-term discounted AoI under different policies}
	\begin{tabular}{c|l|c|c}
		\toprule[2px]
		\textbf{Scenario}&\multicolumn{1}{c|}{\textbf{Policy}}&{$\Delta\overline{D}$}&{$\text{Var}_D$}\\\midrule[1.5px]
		\multirow{4}{*}{\textbf{1}}&\textbf{MDP optimum}	& 1.1808 & $3.6537~\mathrm{e}-3$ \\
		&\textbf{One-step optimum}	&1.2383 & $3.5642~\mathrm{e}-3$\\
		&\textbf{Min error rate}	&1.2397 & $3.0787~\mathrm{e}-3$\\
		&\textbf{Uniform allocation}\footnotemark[1]	&N/A & N/A\\
		\midrule[1px]
		\multirow{4}{*}{\textbf{2}}&\textbf{MDP optimum}	& 0.64029 & $8.4297~\mathrm{e}-4$\\
		&\textbf{One-step optimum}	& 0.68276 & $9.5663~\mathrm{e}-4$\\
		&\textbf{Min error rate}	& 0.68036 & $5.9690~\mathrm{e}-4$\\
		&\textbf{Uniform allocation}\footnotemark[1]	&N/A & N/A\\
		\midrule[1px]
		\multirow{4}{*}{\textbf{3}}&\textbf{MDP optimum}		&$5.9570~\mathrm{e}-3$	&$2.7321~\mathrm{e}-10$\\
		&\textbf{One-step optimum}	&$6.3155~\mathrm{e}-3$		&$1.1603~\mathrm{e}-9$\\
		&\textbf{Min error rate}	&$6.3148~\mathrm{e}-3$			&$3.9019~\mathrm{e}-10$\\
		&\textbf{Uniform allocation}	&0.27194			&$8.0836~\mathrm{e}-5$\\
		\midrule[1px]
		\multirow{4}{*}{\textbf{4}}&\textbf{MDP optimum}	&0.0117			&$6.0254~\mathrm{e}-10$\\
		&\textbf{One-step optimum}	&0.0117			&$1.9271~\mathrm{e}-9$\\
		&\textbf{Min error rate}	&0.0117			&$2.5542~\mathrm{e}-9$\\
		&\textbf{Uniform allocation}	&0.0117			&$1.5954~\mathrm{e}-9$\\
		\bottomrule[2px]
	\end{tabular}
	\label{tab:longterm_performance}
\end{table}
\begin{table}[!tbp]
	\centering
	\caption{Undiscounted AoI under different policies}
	\begin{tabular}{c|l|c|c}
		\toprule[2px]
		\textbf{Scenario}&\multicolumn{1}{c|}{\textbf{Policy}}&{$\Delta\overline{\vert\mathbf{A}\vert}$}&{$\text{Var}_{\vert\mathbf{A}\vert}$}\\\midrule[1.5px]
		\multirow{4}{*}{\textbf{1}}&\textbf{MDP optimum}	& 0.11751 & $1.2484\times10^{-1}$\\
		&\textbf{One-step optimum}	&0.12476 & $1.3137\times10^{-1}$\\
		&\textbf{Min error rate}	&0.12436 & $1.3338\times10^{-1}$\\
		&\textbf{Uniform allocation}\footnotemark[1]	&N/A & N/A\\
		\midrule[1px]
		\multirow{4}{*}{\textbf{2}}&\textbf{MDP optimum}	& $6.4327~\mathrm{e}-2$ & $6.7281~\mathrm{e}-2$\\
		&\textbf{One-step optimum}	& $6.8890~\mathrm{e}-2$ & $7.1667~\mathrm{e}-2$\\
		&\textbf{Min error rate}	& $6.8287~\mathrm{e}-2$ & $7.1265~\mathrm{e}-2$\\
		&\textbf{Uniform allocation}\footnotemark[1]	&N/A & N/A\\
		\midrule[1px]
		\multirow{4}{*}{\textbf{3}}&\textbf{MDP optimum}		&$5.1497~\mathrm{e}-4$	&$5.1464~\mathrm{e}-4$\\
		&\textbf{One-step optimum}	&$6.3155~\mathrm{e}-4$		&$6.8602~\mathrm{e}-4$\\
		&\textbf{Min error rate}	&$6.3148~\mathrm{e}-4$			&$5.9435~\mathrm{e}-4$\\
		&\textbf{Uniform allocation}	&$2.7481~\mathrm{e}-2$			&$2.8254~\mathrm{e}-2$\\
		\midrule[1px]
		\multirow{4}{*}{\textbf{4}}&\textbf{MDP optimum}	&$1.1856~\mathrm{e}-3$	&$1.1843~\mathrm{e}-3$\\
		&\textbf{One-step optimum}	&$1.3174~\mathrm{e}-3$		&$1.1316~\mathrm{e}-3$\\
		&\textbf{Min error rate}	&$1.1058~\mathrm{e}-3$			&$1.1047~\mathrm{e}-3$\\
		&\textbf{Uniform allocation}	&$1.1816~\mathrm{e}-3$			&$1.1880~\mathrm{e}-3$\\
		\bottomrule[2px]
	\end{tabular}
	\label{tab:instant_performance}
\end{table}
\footnotetext[1]{Sensor 1 fails to deliver the required minimal packet transmission rate $(1-\varepsilon_{\max})$ under the uniform allocation policy in Scenarios 1 \& 2.}

\section{Further Discussions}\label{sec:discussions}
\subsection{Fading Channels}
In this study we have assumed non-fading channels for simplification so far.  In practical applications, the random fluctuation of channel conditions must be taken into account. In this case, the terms $\varepsilon_1(\mathbf{A}_{k-1})$ and $\varepsilon_2(\mathbf{A}_{k-1})$ in \eqref{eq:one-step-aoi-min} and \eqref{eq:transition} must be replaced by their expectations $\mathbb{E}\left\{\varepsilon_1\right\}$ and $\mathbb{E}\left\{\varepsilon_2\right\}$, respectively. Nevertheless, as long as the CSI is available at the cloud server or measurable with sufficient accuracy, both the expectations can be straightforwardly calculated without impacting the deployment of our proposed methods.

\subsection{Computational Complexity of the MDP Method}
To obtain the optimal $\mathbf{n}$ from a trained Q-matrix according to \eqref{eq:mdp_opt_policy} is a straightforward inquiry from indexed data set, of which the computational effort is negligible. However, the iterative Q-learning process to train the Q-matrix according to Algorithm~\ref{alg:rl_solver} has a time complexity of $\mathcal{O}(IJl_{\max})$, where in the $M$-sensor case $I=A_{\max}^M$ and $J\sim N_{\max}^M$. This leads to a dramatic increase of learning effort as $M$ increases to a common level in practical sensor networks, which becomes a huge challenge for the deployment of the proposed long-term AoI optimization method, especially when taking the long-term inconsistency of channels into account. Fast solvers for the MDP \eqref{eq:mdp_finite} are therefore required. Possible approaches to boost the MDP solution include fast online learning, clustering sensors with similar CSI and applying heuristic algorithms.

\subsection{Multi-Hop Clustering}
In context of massive Machine-Type Communications (mMTC) scenarios such as Internet-of-Things (IoT) and sensor networks, our previous works have encouraged to deploy a multi-hop architecture, where devices are grouped into multiple clusters, each with a head that relays messages for other cluster members~\cite{Han_2017, Ji_2017}. It has been demonstrated that appropriate clustering and head selection are critical to control the message delay in uplink. Obviously, the AoI at the server will also be significantly impacted by the same factors, which is worth further study.

\section{Conclusion}\label{sec:conclusion}
In this paper, we have pioneered to bridge the gap between two significant research areas: the Age of Information and the finite blocklength information theory. Our study begins with a novel AoI model for sensor networks, in which the system AoI is formulated as a function of sensors' transmission error rates. By linking the error rates to the blocklength assignment we are able to propose to optimize the system in terms of both instantaneous undiscounted AoI and long-term discounted AoI. While the former one can be solved in a straight-forward manner, the latter one is an infinite-state MDP that can be approximated to a finite-state version with good accuracy, and solved with RL technique. Our simulations have demonstrated that the long-term optimization method outperforms all other methods, while the one-step optimizer fails, as we have expected, to deliver a significant performance gain in comparison to the error-rate oriented benchmark. We have also provided several potential extensions of this work for future study, each supplemented by a brief discussion.

\section*{Acknowledgment}
This work has been supported by the \emph{Federal Ministry of Education and Research (BMBF) of the Federal Republic of Germany}, in scope of the project ``5GANG'' (Grant Number 16KIS0725K). The authors alone are responsible for the content of this paper.

\ifCLASSOPTIONcaptionsoff
  \newpage
\fi

%
%
%
%
%




\end{document}